\documentclass[journal]{IEEEtran}
\usepackage{nopageno}
\usepackage{romannum}
\usepackage[usenames,dvipsnames]{xcolor}
\usepackage{tkz-berge}
\usetikzlibrary{fit,shapes}                                     
\usepackage{calrsfs}
\usepackage{graphicx}
\usepackage{graphics} 
\usepackage{epsfig}    
\usepackage{mathptmx}
\usepackage{caption}
\usepackage{subcaption}
\usepackage{times} 
\usepackage{tikz}
\usetikzlibrary{positioning}
\usepackage{amsmath, amssymb}
\usetikzlibrary{arrows}

\usepackage{enumitem,kantlipsum}
\usepackage{amsthm}
\usepackage{amsfonts} 
\usepackage{setspace}
\usepackage{multirow}
\usepackage{textcomp}
\usepackage[english]{babel}
\usepackage{float} 
\usepackage{algorithmicx}
\usepackage[section]{algorithm}
\usepackage{algpascal}
\usepackage{algc}
\usepackage{algcompatible}
\usepackage{algpseudocode}
\let\oldReturn\Return
\renewcommand{\Return}{\State\oldReturn}
\usepackage{mathtools}
\usepackage{bm}
\usepackage{mathptmx}
\usepackage{times} 
\usepackage{mathtools, cuted}
\usepackage{textcomp}
\usepackage[english]{babel}
\usepackage{rotating}
\usepackage{pgfplots}
\pgfplotsset{compat=1.5}
\usepackage{siunitx}
\usepackage{pgfplotstable}
\usepackage{filecontents}

\def\BibTeX{{\rm B\kern-.05em{\sc i\kern-.025em b}\kern-.08em
    T\kern-.1667em\lower.7ex\hbox{E}\kern-.125emX}}

\begin{document}
\pagestyle{empty}
\title{MIST: A Novel Training Strategy for Low-latency Scalable Neural Net Decoders} 
\author{Kumar Yashashwi\textsuperscript{*}, Deepak Anand\textsuperscript{*}, Sibi Raj B Pillai\textsuperscript{*},
Prasanna Chaporkar\textsuperscript{*}, K Ganesh\textsuperscript{$\dagger$}\\ 
\textsuperscript{*}Department of Electrical Engineering, Indian Institute of Technology Bombay, India \\  \textsuperscript{$\dagger$}Manufacturing \& Supply Chain Center of Competence, McKinsey \& Company, India\\
Email: \textsuperscript{*}\{kryashashwi, deepakanand,  bsraj, chaporkar\}@ee.iitb.ac.in, \textsuperscript{$\dagger$}k\_ganesh@mckinsey.com}


\maketitle
\thispagestyle{empty}
\begin{abstract}
In this paper, we propose a low latency, robust and scalable neural net based decoder for convolutional and low-density parity-check (LPDC) coding schemes. The proposed decoders are demonstrated to have  bit error rate (BER) and block error rate (BLER) performances at par with the state-of-the-art neural net based decoders while achieving more than 8 times higher decoding speed. The enhanced decoding speed is due to the use of convolutional neural network (CNN) as opposed to recurrent neural network (RNN) used in the best known neural net based decoders. This contradicts existing doctrine that only RNN based decoders can provide a performance close to the optimal ones. The key ingredient to our approach is a novel Mixed-SNR Independent Samples based Training (MIST), which allows for training of CNN with only 1\% of possible datawords, even for block length as high as 1000. The proposed decoder is robust as, once trained, the same decoder can be used for a wide range of SNR values. Finally, in the presence of channel outages, the proposed decoders outperform the best known decoders, {\it viz.} unquantized Viterbi decoder for convolutional code, and belief propagation for LDPC. This gives the CNN decoder a significant advantage in 5G millimeter wave systems, where channel outages are prevalent.
\end{abstract}
\vspace{-0.1cm}
\section{Introduction}\label{intro}
%
Efficient communication of messages over a noisy channel is governed by information theoretic principles. While communication efficiency, measured in terms of the successful transmission rate, can be optimized using  careful code design, computational efficiency is also of at most importance in modern wireless devices. In fact
thoughtful design of encoding and decoding schemes such as \emph{convolutional, turbo, LDPC and polar} have successively pushed the operational rates closer to the maximum possible limit, known as Shannon capacity of the channel~\cite{lin2001error}. In conjunction with iterative decoding, the computational complexity also stays within manageable limits, making these schemes good candidates for current and future wireless standards. Recently, new decoding
schemes based on deep-learning have shown to have good performance for decoding \emph{polar} codes~\cite{gruber2017deep}. This was later extended to \emph{turbo} and \emph{convolutional} codes~\cite{kim2018communication}.

The most attractive feature of a learning based decoder is that it provides a kind of universality to the decoding scheme, i.e. the same network architecture can be properly trained to decode on a variety of channels. 
This leads to a certain ease of implementation, which along with parallel processing can also bring
hardware advantages. Another significant advantage of learning based methods over conventional decoding approaches is that they can be made more agnostic to the exact statistics of the underlying channel. More specifically, while variations from the designed system parameters can be harmful for even the optimal MAP decoder, a learning based decoder is more robust against channel variations~\cite{farsad2018neural}. 
In other words, the network can be trained to guard against channel variations also, albeit at
the expense of some training cost, an one time expenditure. 
In communication systems, deep learning has also been shown to be highly useful in tasks such as estimating channel state information (CSI)~\cite{ye2018power}, noise parameter estimation~\cite{liang2018iterative} and modulation recognition~\cite{yashashwi2019learnable}. Designing neural net based decoder for convolutional and LDPC coding schemes is the main aim of the current paper. 




Some references to  the prior work is in order here. In~\cite{o2017introduction}, a deep learning based auto-encoder strategy for designing an end-to-end communication system was proposed for low block lengths. 
In fact an error correction performance comparable to that of $(7,4,3)$ Hamming code was demonstrated, under suitable assumptions on the channel model. \cite{liang2018iterative} proposes a CNN based method to estimate noise parameters and use it in tandem with belief propagation decoder to reduce the decoding error rates for LDPC codes. Short BCH codes are decoded using RNN based methods  in~\cite{nachmani2018deep}.
In~\cite{gruber2017deep}, a Multi Layer Perceptron (MLP)
based method for decoding polar codes is proposed. This work is extended in~\cite{lyu2018performance}, where decoders for polar codes using CNN and RNN
are proposed, and their accuracy and time complexity are compared with MLP. It is shown that though RNN based decoder has worse time complexity, it achieves the lowest error rates among other deep learning  based methods. This work also argues that the achieved performance can be
close to optimal only when at least 90\% of the codewords are used for training. 
Furthermore, the CNN decoding performance was shown to be inferior even with this disproportionate training requirement. Clearly, such training constraints are only feasible for low blocklengths like $32$.
In their seminal paper, Kim {\it et al.} have
proposed RNN based decoder for recursive convolutional codes~\cite{kim2018communication}. 
The proposed decoder achieves close to optimal performance and can also scale
to higher block lengths. The training  was done using a sufficiently large, yet computationally
feasible set of input codewords. In spite of these desirable characteristics,
the decoder in~\cite{kim2018communication} has high decoding latency
that limits its use in practice. While CNNs are known to have low computational latency, their
performance so far was perceived poor when compared to the RNNs. The main question is 
whether one can realize a CNN based decoder which can perform on par with RNN decoders, at comparable blocklengths. 
In this paper, we show that the key to building such a CNN is to employ an efficient training mechanism. In fact, our novel training scheme in conjunction with a better network architecture allows for building low latency CNN decoders for both convolutional and LDPC codes. 

The key contributions of the paper are:

\begin{enumerate}[wide, labelwidth=!, labelindent=0pt]

\item[$\bullet$] A novel Mixed-SNR Independent Samples based Training (MIST) scheme is proposed. With this training scheme, it is shown that deep learning based CNN decoders for
convolutional codes can be designed, achieving better BER and  BLER performance than that of a hard Viterbi decoder. 
\item[$\bullet$] CNN based decoders with MIST are shown to achieve similar performance as the state-of-the-art RNN based decoders for convolutional codes, that too
with less than 1\% of the codebook
used for training, as against the existing 90\% training
requirement~\cite{lyu2018performance}.
\item[$\bullet$] The {\it same} neural net architecture can be trained using MIST to decode LDPC codes. In fact this decoder achieves a better performance than that of the popular bit flipping algorithm~\cite{lin2001error}.
\item[$\bullet$] The proposed method achieves a lower BER compared to both hard and soft decoding techniques for convolutional and LDPC codes when there are random outages in a channel. 
The issue of random outages are expected to play a key role in 5G  systems~\cite{rappaport2013millimeter}.
\item[$\bullet$] Finally, MIST procedure scales well with blocklength. In our experiments using comparable blocklengths, it achieves $8$ times lower latency than existing state-of-the-art neural net methods, while guaranteeing the same
BER and BLER performance.
\end{enumerate}

Rest of the paper is organized as follows: 
Section~\ref{sec:method} describes the proposed training strategy and neural net architecture. Section~\ref{sec:results} compares the performance of the proposed method with the RNN decoder and traditional decoding methods. Section~\ref{sec:conclusion} gives concluding remarks. 

\vspace{-0.2cm}
\section{System Model}

\begin{figure}[t]
\begin{tikzpicture}[rect/.style={minimum height=1cm, text width=1.25cm, text centered, rectangle, draw}, line width=1.5pt, scale=0.92]
\node (enc) at (0,0)[rect] {Encoder};
\node (mod) at (2.25,0)[rect]{BPSK};
\node (cha) at (4.5,0)[rect]{Channel};
\node (dec) at (6.75,0) [rect]{Neural Net};
\draw[->] (enc) -- node[above]{$x$} (mod);
\draw[->] (mod) -- node[above]{$b$} (cha);
\draw[->] (cha) -- node[above]{$y$}(dec);
\draw[<-] (enc) --++(-1.25,0) node[below]{$m$};
\draw[->] (dec) --++(1.25,0) node[below]{$\hat m$};
\end{tikzpicture}

\caption{\label{fig:systemmodel} \small System model}  
\end{figure}
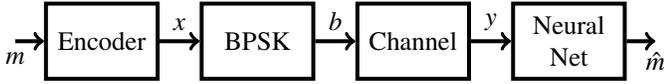

The problem of designing an optimal decoder can be formulated as follows.
Consider a typical communication system  shown in 
Figure~\ref{fig:systemmodel}, where an $l-$bit dataword $\mathbf{m} \in \{0,1\}^l$ is encoded
to $n-$length codeword $\mathbf x\in \{0,1\}^n$. Here $n$ is the blocklength, and $r=\frac ln$ is
called the rate of the code.
 Let $\bm{m} \in  \mathcal{M}$, where $\mathcal{M} = \{0,1\}^\ell$. The chosen codeword $\bm{x}$ is modulated 
 to yield the baseband transmitted symbols $\bm{b}$. In this paper we assume binary phase shift keying (BPSK) as the digital modulation scheme. 
 The received baseband waveform after sampling yields the observation vector $\bm{y}$. We
 consider an AWGN model in which $\bm{y}$ is given by
 \begin{align} \label{eq:model}
 \bm{y} = \bm{b} + \bm{z},
 \end{align}
 where $\bm z$ is a real zero-mean Gaussian noise vector of covariance matrix $\sigma^2 \mathbb I_n$, 
 independent of the transmitted symbols. 
 While the above model depicts a block-coding scheme, with slight abuse of terminology, this
 also applies to non-block coding schemes like convolutional codes. Note that the same system model is considered in previous work on deep learning based decoder design~\cite{gruber2017deep},~\cite{kim2018communication},~\cite{lyu2018performance}.
 The decoder's goal now is to find a mapping $f^{\star}: \Re^{n} \rightarrow {\cal M}$ such that:
\begin{align}
f^{\star}(\mathbf{y})=\arg\max_{\!\!\!\!\!\!\!\!\!\!\mathbf{m} \in \mathcal{M}}~\mathbb{P}(\bm{m}|\bm{y}).
     \label{eq:map}
\end{align}
A learning based decoder effectively learns the function $f^{\star}$ during the training stage, typically using labelled data comprising  input codewords and the corresponding noisy received vectors. Next, we describe the proposed method.

\vspace{-0.3cm}
\section{Proposed Method}\label{sec:method}
In this section, we explain our training procedure, neural net design, choice of hyperparameters, and the optimization performed for tuning weights of the neural net. Recall that our objective is to decode convolutional and LDPC codes. The training procedure plays a key role in determining the performance of a neural net. For most classification problems, conventional training methods partition the given labelled data into three sets, viz. 1)~{\it training} dataset, 2)~{\it validation} dataset and 3)~{\it testing} dataset. The training dataset is first used to train the network, i.e., weights in the neural net are tuned to minimize an appropriately chosen \emph{loss function}. Subsequently, validation data is used to detect over-fitting, and following that the hyperparameters of the network are tuned to eliminate the over-fitting. The training and validation steps are typically iterated until the hyperparameters are tuned to yield the desired level of accuracy.
Once the neural net is optimized, the results are shown on the testing data. 

Note that decoding is effectively a classification problem, thus the three step methodology  described above is universally employed for decoding schemes also (see~\cite{kim2018communication} and~\cite{farsad2018neural} for details). However, 
for a typical classification problem,  training is done using a fixed labelled 
dataset, as obtaining more data is often expensive. On the other hand, for a given
encoding rule and channel probability law, the training data for a decoder can be generated easily. 
We leverage this to propose a training procedure that generate labelled training
data {\it on-the-fly}. 
Thus, unlike the classification framework which employs a fixed fraction of the codewords in training, we randomize the set of inputs during the training stages, yielding significant advantages while decoding.
Another key difference in our training methodology is the way in which 
we account for various SNR values encountered by the decoder. Note that the previous works use an
appropriately chosen SNR for training, and this does not change during the training process.
In contrast, the proposed training method randomly samples SNR value from a desired range for each training codeword.
We call our training strategy as {\it Mixed-SNR Independent Sampling based Training} (MIST), which is described  next.

Before getting into further details, it is worthwhile to note that extensive simulations were
conducted before a satisfactory CNN architecture emerged, and the two above-mentioned essential training features of MIST give dramatic performance improvements  from conventionally trained CNNs.

\subsection{Mixed-SNR Independent Sampling based Training (MIST)}
Let ${\cal S}$ be the set of SNR values for which a decoder has to be designed. 
The training is done using batches of valid codewords generated for the coding scheme under consideration. Here, we are interested in convolutional and LDPC codes. 
The following steps are performed $\beta$ times to obtain a training batch with $\beta$ codewords. 
\begin{enumerate}[wide, labelwidth=!, labelindent=0pt]
\item[$\bullet$] Uniformly sample a SNR value from ${\cal S}$ 
and set the noise variance as $SNR^{-1}$. 
The noise vector $\bm{z}$ is then found by independently sampling the noise distribution $n$ times.
\item[$\bullet$] Sample a message $\bm{m}$ from $\mathcal{M}$, generate the codeword $\bm x (\bm m)$,
and output symbols $\bm{y}$ using \eqref{eq:model}.
\end{enumerate}
Let the training batch be denoted as $\mathbf{y}_{1},\mathbf{y}_{2},\ldots,\mathbf{y}_{\beta}$, 
having dimensions $\beta \times n$. The training batch is feedforwarded to the neural net that 
outputs a $\ell$-dimensional vector $\bm{\hat p}$. The mean squared error (MSE) between $\bm{\hat p}$ and $\bm{m}$ is calculated as: 
\begin{align}
L(\bm{m},\bm{\hat p}) =\dfrac{1}{\beta\ell}\sum_{i=1}^{\beta}\sum_{j=1}^{\ell} (m_{ij} - {\hat p}_{ij})^2,
     \label{eq:mse}
\end{align}
where
$m_{ij}$ (${\hat p}_{ij}$, resp.) denotes $j^{\rm th}$ value in vector $\bm{m}_i$ 
($\bm{\hat p}_i$, resp.). The function $L(\cdot,\cdot)$ in equation~(\ref{eq:mse}) is our {\it loss function}, and it is back-propagated to learn the weights of the neural net using the {\em Adam optimizer} \cite{gruber2017deep}.  Choosing a sufficiently small learning rate for the CNN training is now 
important, however, too small a value will delay the convergence of the learning algorithm. 
Experimental validation suggests a suitable initial learning rate of $10^{-3}$. 
A new training batch is generated in each iteration.

The key differences between the existing training strategies and MIST are as follows:
\begin{enumerate}[wide, labelwidth=!, labelindent=0pt]
\item[$(a)$]Existing works fix the training and validation datawords in the beginning, 
and use the same in every training iteration. On the contrary,   
MIST generates the training data randomly using Monte-Carlo technique in every training
iteration. 
\item[$(b)$]MIST does not require any separate validation data for hyperparameter tuning
as over-fitting is highly improbable. Rather, as explained in Section~\ref{sec:method_architecture}, hyperparameters are tuned directly based on the value of loss function.
\item[$(c)$]Existing works
train the network for a single suitable SNR value, and then tests on different SNR values.
In MIST, each sample in a batch is allowed to have randomly chosen SNR from the given set ${\cal S}$.
\end{enumerate}

Some advantages of MIST are: 
\begin{enumerate}[wide, labelwidth=!, labelindent=0pt]
\item[(1)]
Since the storage requirements of the training data is low in MIST, the network can be trained on a larger set of datawords even for a large block lengths.
\item[(2)]As the training samples vary across training iterations, the neural
net actually learns the decoding technique 
instead of becoming a simple 
recording and reading algorithm, thus avoiding an overfit to the training data. 
\item[(3)]As the noise
is varied across samples, the neural net gets to see the samples 
corresponding to various SNRs. This helps it tune the decoding function
to accommodate varying SNR. 
\item[(4)]MIST
allows for a low complexity neural net to perform close to optimal with
significantly less decoding time per message than that of the existing
neural net based approaches.
\end{enumerate}

Next we describe the neural net architecture used for decoding.

\vspace{-0.1cm}
\subsection{Network Architecture}\label{sec:method_architecture}
\begin{figure}

\centering
\includegraphics[width=3in]{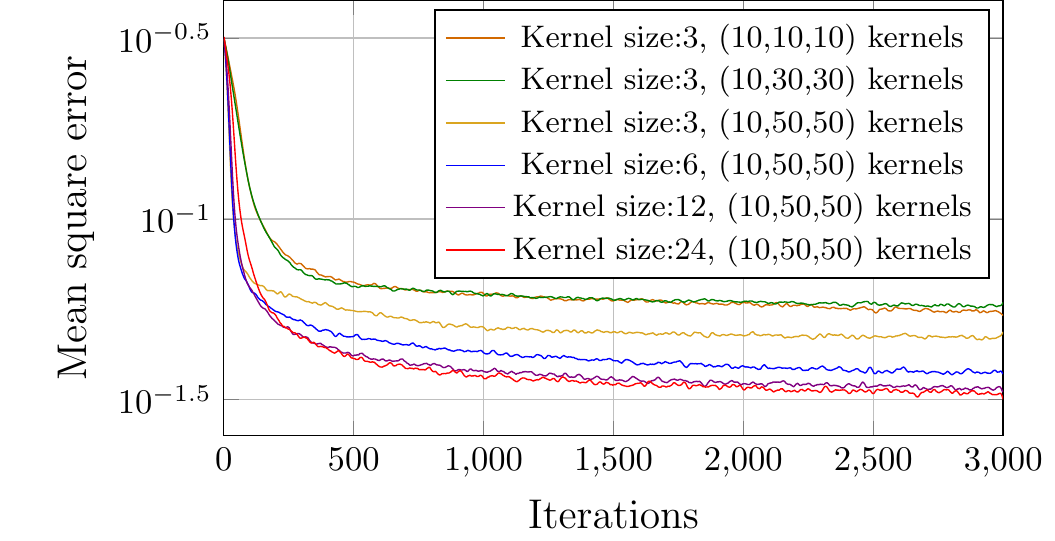} 
\caption{\label{fig:training} \small Training procedures to decide on hyperparameters for $3$ layer CNN. Each entry in legend denotes the kernel size used in all the $3$ layers followed by the number of kernels in each layer.}  
\end{figure}
Among different deep learning based approaches, we choose CNNs. One-dimensional CNN kernels are capable of learning the decoding algorithm as it models the {\it sense of sequence} necessary for learning dependencies among encoded message bits. Moreover CNNs have lower training and testing complexity, which are highly desired. The architecture of a CNN is decided by $3$ hyperparameters, \emph{viz.}  1)~number of layers, 2)~number of kernels in each layer,  3)~kernel size in each layer. We reiterate that extensive computational experiments led to the final working architecture and hyperparameters presented below. 


Having a network architecture with $2$ layers
was found insufficient to achieve reasonable performance in most of our experiments. While this behaviour dramatically changed with $3$ layers, our experiments further showed that increasing layers beyond $3$ did not improve the performance by much. In order to choose the number of kernels,
we started with lower values and systematically increased the number till reasonable performance 
was achieved. This is depicted in Figure~\ref{fig:training}, where the error performance
for various kernel choices are listed. While choosing about $50$ kernels for the last two layers
gave reasonable results, the first layer could operate with just $10$ kernels. 
Increasing the number of kernels beyond $50$ did not provide much improvement in the error rate.
These comparisons were done for a kernel size of $3$.
%
Once the number of kernels are fixed, the kernel size is determined. Since a codeword has a fixed bit sequence, a larger sequence context is required by the CNN to learn the decoding scheme. Thus using a larger kernel size can help reducing the error  further, this is illustrated in Figure~\ref{fig:training}. 
Notice that when the kernel size is $3$, the error is higher compared to the case when kernel size is $6$. While increasing the kernel size to $12$ and then $24$  decreases the error further, no noticeable improvement was obtained after this. The activation function used in each convolutional layer is rectified linear unit (ReLU). 

The convolution was performed by padding with zeros at signal edges. Since MIST ensures no overfit, \emph{dropouts} are not used while training the deep network. It is experimentally verified that adding dropout does not reduce the loss function any further. Once the signal features are extracted from the convolutional layer, it is passed through the dense layer having $\ell$ neurons with sigmoid activation to get the posterior probability corresponding to each message bit. 
This posterior probability is quantized to obtain the message bits.
The experimentally obtained CNN architecture is given in Table~\ref{table:network} \footnote{Code available at https://github.com/kryashashwi/MIST\_CNN\_Decoder}. The same neural net design is used for decoding both convolutional as well as LDPC codes.

\begin{table}
\centering
\caption{\small Network architecture for the proposed decoders}
\label{table:network}
\begin{tabular}{|c|c|c|}
\hline
\textbf{Layer} & \textbf{Output shape} \\

\hline
Conv1 (ReLU) & ($\beta, n, 10$)\\
\hline
Batch normalization & ($\beta, n, 10$)\\
\hline
Conv2 (ReLU)  & ($\beta, n, 50$)\\
\hline
Batch normalization & ($\beta, n, 50$)\\
\hline
Conv3 (ReLU)& ($\beta, n, 50$)\\
\hline
Batch normalization & ($\beta, n, 50$)\\
\hline
Dense (Sigmoid) & ($\beta, l$)\\
\hline
\end{tabular}
\end{table}

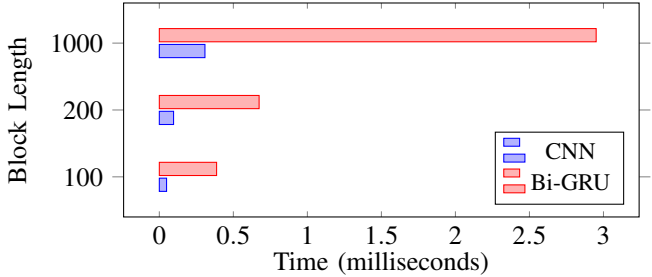
\begin{figure}
\centering
\begin{tikzpicture}
\begin{axis}[xbar, symbolic y coords={100,200,1000}, ytick=data, yscale=0.5, 
enlarge y limits=0.3, ylabel={Block Length},legend style={{font=\small},at={(0.72,0.61)},anchor=north west},
xlabel={Time (milliseconds)}, xlabel shift=0.2cm]
\addplot
coordinates{(0.048987716,100) (0.096297473,200) (0.308185875,1000)};

\addplot
coordinates{(0.38692929,100) (0.67403546,200)(2.95073688,1000)};
\legend{CNN,Bi-GRU}
\end{axis}
\end{tikzpicture}
\caption{\label{fig:time} \small Average time taken to decode a dataword using CNN decoder and Bi-GRU decoder for blocklength $100$, $200$ and $1000$.}  
\end{figure}

\vspace{-0.1cm}
\section{Results and Discussion}\label{sec:results}
We present the BER and BLER curves for the  CNN decoder in AWGN channels, against the available SNR.
The performance is evaluated on $10^{9}$ samples for each SNR value.
Comparison with analytically derived decoders is performed for both convolutional and LDPC codes. 

\vspace{-0.1cm}
\subsection{Decoding of convolutional encoding method}
\begin{figure*}[ht]
   \begin{subfigure}[!htb]{0.32\textwidth}
   \centering
    \includegraphics[width=2.35in, height=1.72in]{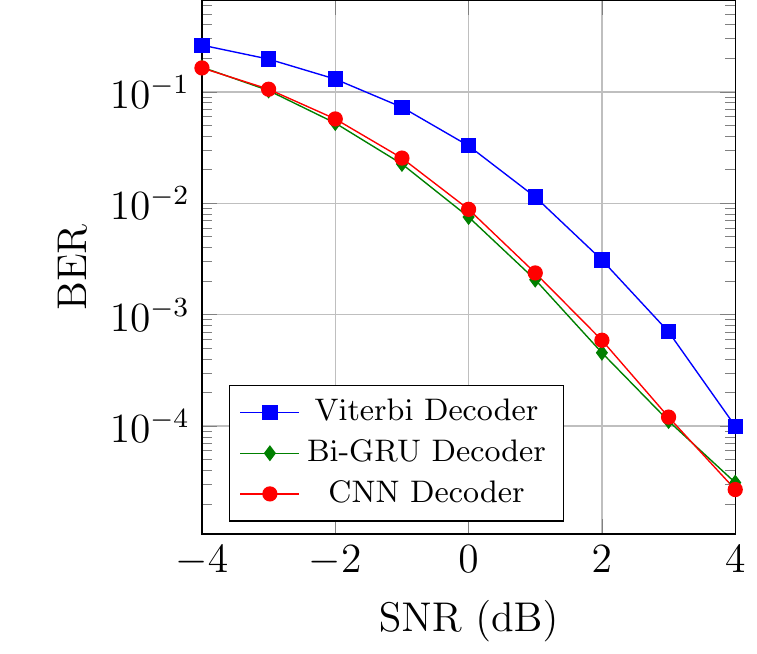}
    \caption{BER Vs SNR ($n$=100)}
    \label{fig:50ber}
    \end{subfigure}
     ~
    \begin{subfigure}{0.32\textwidth}
    \centering
    \includegraphics[width=2.35in, height=1.72in]{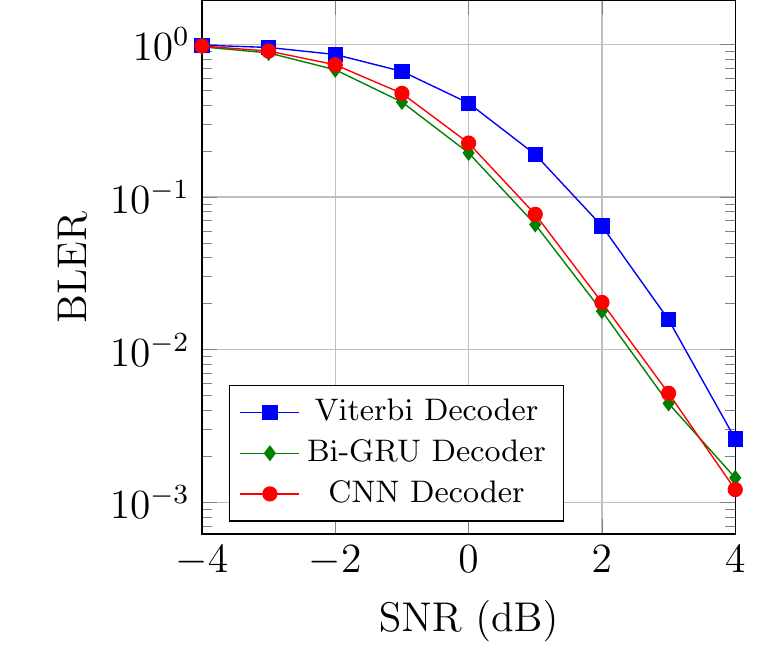} 
    \caption{BLER Vs SNR ($n$=100)}  
    \label{fig:50bler}
    \end{subfigure}
    ~
    \begin{subfigure}[!htb]{0.32\textwidth}
   \centering
    \includegraphics[width=2.35in, height=1.72in]{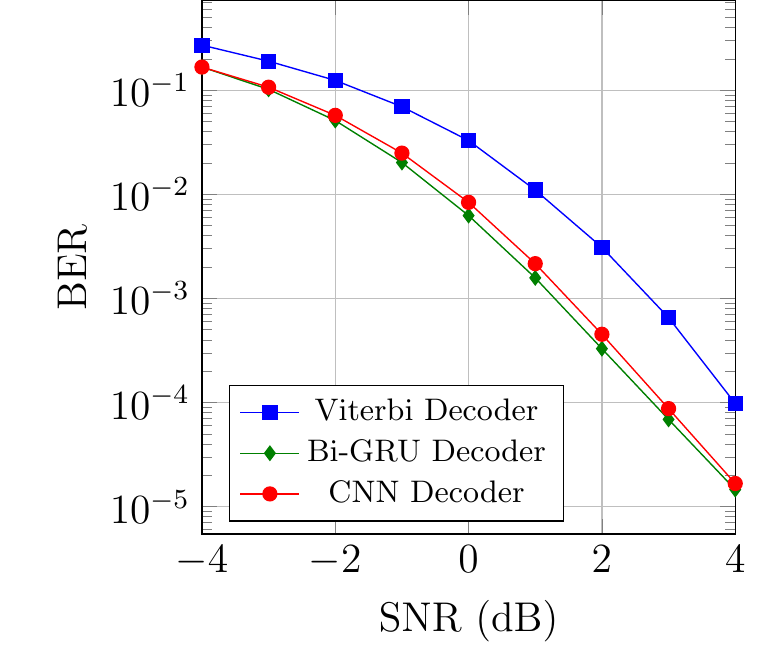} 
    \caption{BER Vs SNR ($n$=200)}
    \label{fig:100ber}
    \end{subfigure}
     ~
    \begin{subfigure}{0.32\textwidth}
    \centering
    \includegraphics[width=2.35in, height=1.72in]{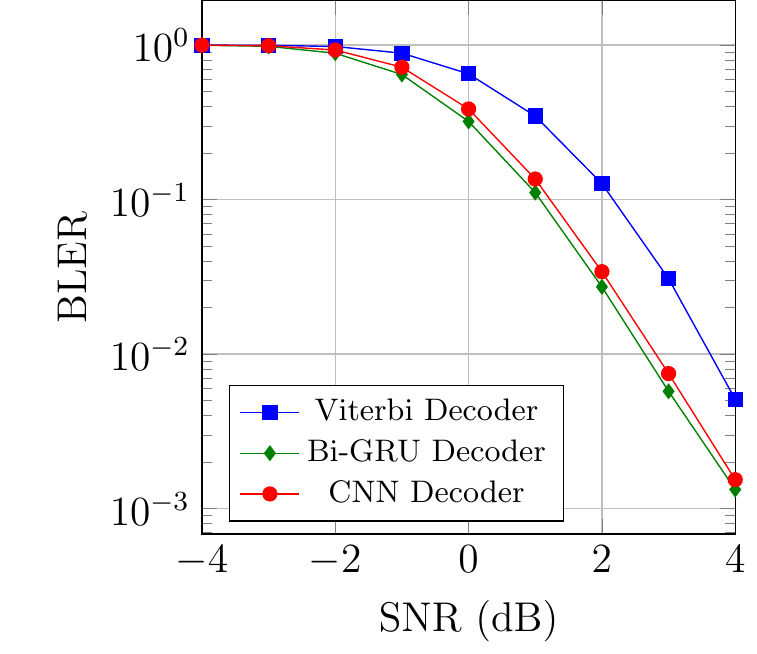} 
    \caption{BLER Vs SNR ($n$=200)}  
    \label{fig:100bler}
    \end{subfigure}
     ~
    \begin{subfigure}{0.32\textwidth}
    \centering
    \includegraphics[width=2.35in, height=1.72in]{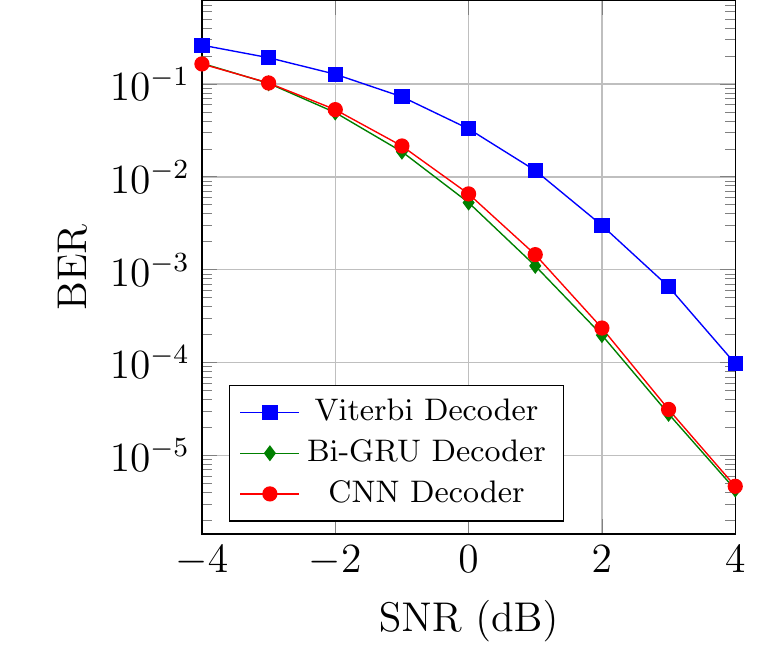} 
    \caption{BER Vs SNR ($n$=1000)}  
    \label{fig:500ber}
    \end{subfigure}
     ~
    \begin{subfigure}{0.32\textwidth}
    \centering
    \includegraphics[width=2.35in, height=1.72in]{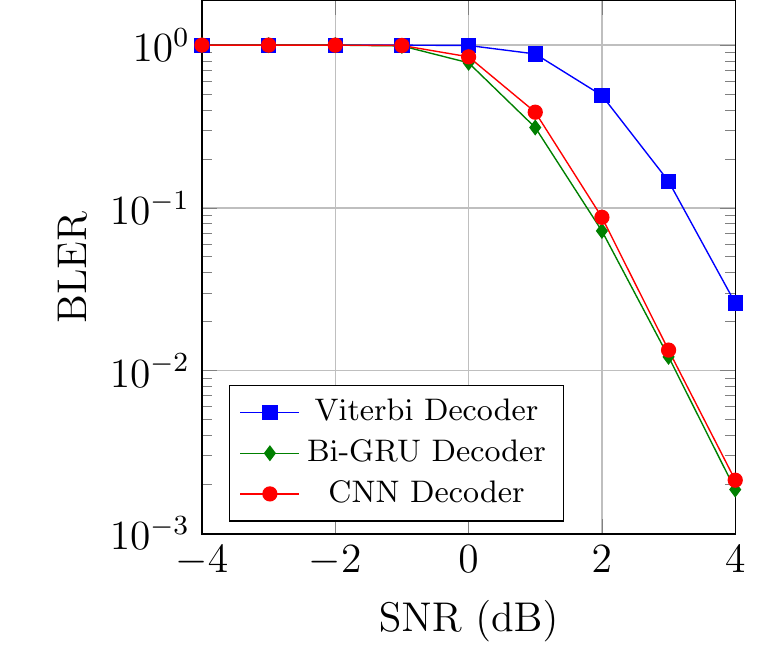} 
    \caption{BLER Vs SNR ($n$=1000)}  
    \label{fig:500bler}
    \end{subfigure}
\caption{\label{fig:convolutionalEncoder} BER and BLER comparison of Hard Viterbi decoder, Bi-GRU decoder and CNN decoder.
}
\end{figure*}

Generating polynomials for convolutional codes with good minimum distance properties are readily
available~\cite{lin2001error}. For illustration, the performance on a non-systematic rate-$\frac 12$ code is compared against
a hard Viterbi decoder (same code and comparison as in \cite{kim2018communication}). The code polynomials in octal notation are $(5,7)$. The minimum free distance is $5$ for this code~\cite{lin2001error}. In Figure~\ref{fig:convolutionalEncoder}, comparisons 
using block lengths $100$, $200$ and $1000$ for  different SNR values is shown. Here, the CNN decoder is compared with the state-of-the-art RNN-based method 
proposed in~\cite{kim2018communication}. The architecture 
in~\cite{kim2018communication} uses $2$ bi-direction Gated Recurrent Units (bi-GRU) layers with batch normalization followed by a final dense layer with sigmoid activation. So far the general perception in literature is that
recurrent layers outperform CNN based decoders, see
\cite{kim2018communication,farsad2018neural,lyu2018performance}.
However, as evident from  Figure~\ref{fig:convolutionalEncoder}, the training strategy MIST with a carefully designed CNN architecture achieves the same performance as the RNN based method. Moreover, CNN decoders have nearly~$8$ times less decoding latency when compared to RNN decoders, as shown in Fig.~\ref{fig:time} \footnote{Time comparison made on machine having Ubuntu~$16.04$, Intel~$8$~core~i$7-7700k@4.2$~GHz, $64$~GB~RAM, Nvidia~GeForce~GTX~Titan-X~$12$~GB~GPU.}.
The time to decode a codeword becomes extremely important when comparing different decoders. To understand the impact of findings in Fig.~\ref{fig:time}, note
that for $n=1000$, a processor running a single thread can maintain a speed of about $170$Kbps while employing a Bi-GRU decoder, while a similar processor will achieve a rate of $1.25$Mbps using the proposed decoder.
{\it Thus, the proposed  CNN based decoder performs on par with the state-of-the-art complex RNN based decoders in terms of BER and BLER, albeit at a much lower latency.}

Now that the usefulness of the proposed training method and network architecture is established, next we show the performance of the CNN decoder for LDPC codes.

\begin{figure*}[!h]
  \begin{subfigure}[!htb]{0.32\textwidth}
  \centering
    \includegraphics[width=2.35in, height=1.72in]{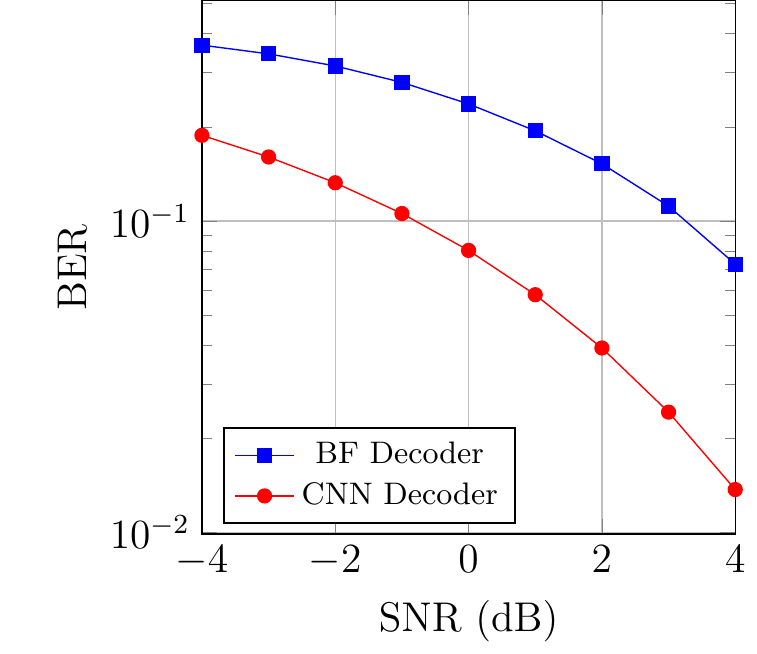} 
    \caption{BER Vs SNR ($n$=100)}
    \label{fig:ldpc50ber}
    \end{subfigure}
     ~
    \begin{subfigure}{0.32\textwidth}
    \centering
    \includegraphics[width=2.35in, height=1.72in]{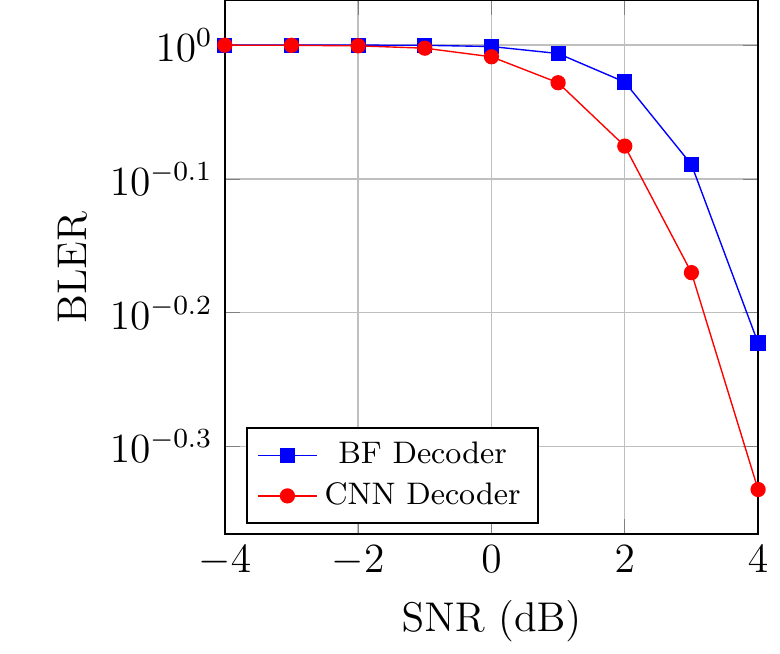}
    \caption{BLER Vs SNR ($n$=100)}  
    \label{fig:ldpc50bler}
    \end{subfigure}
    ~
    \begin{subfigure}[!htb]{0.32\textwidth}
  \centering
    \includegraphics[width=2.35in, height=1.72in]{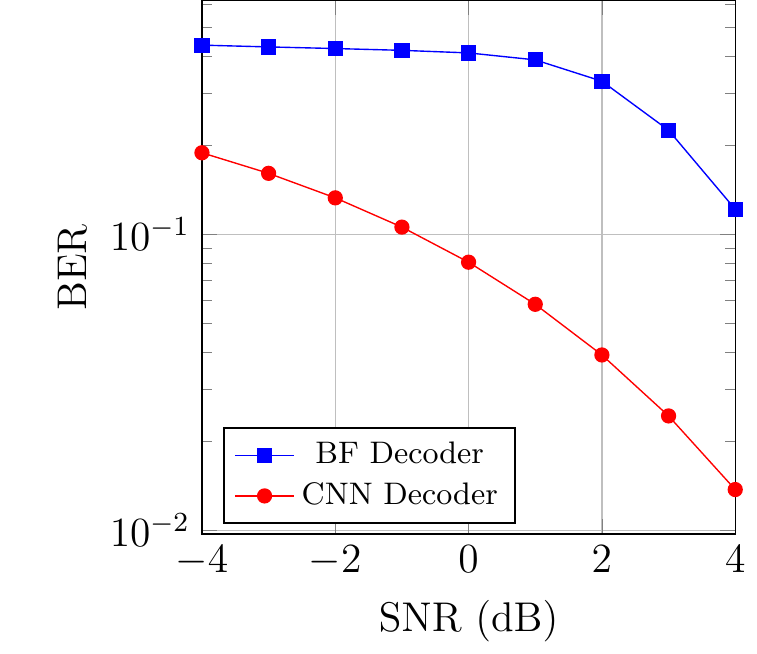} 
    \caption{BER Vs SNR ($n$=200)}
    \label{fig:ldpc100ber}
    \end{subfigure}
     ~
    \begin{subfigure}{0.32\textwidth}
    \centering
    \includegraphics[width=2.35in, height=1.72in]{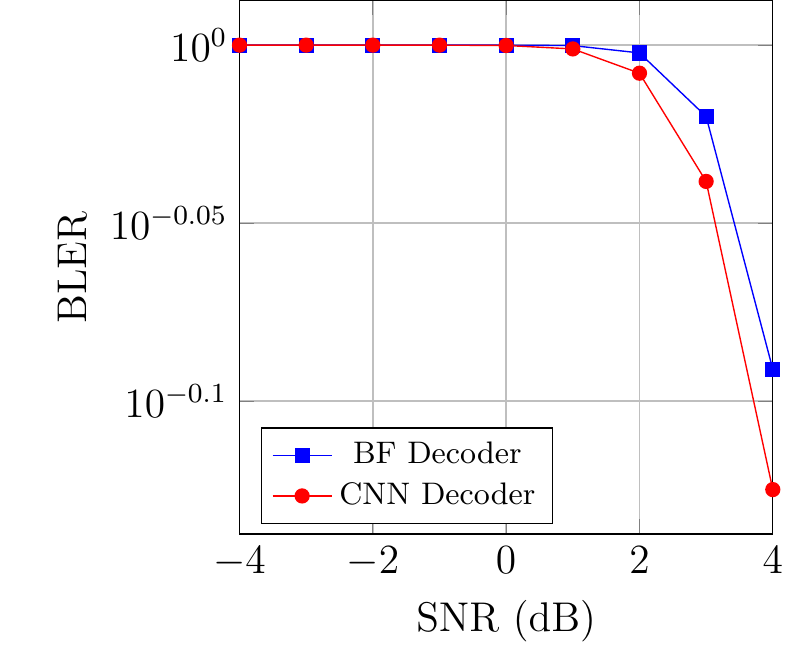}
    \caption{BLER Vs SNR ($n$=200)}  
    \label{fig:ldpc100bler}
    \end{subfigure}
     ~
    \begin{subfigure}{0.32\textwidth}
    \centering
    \includegraphics[width=2.35in, height=1.72in]{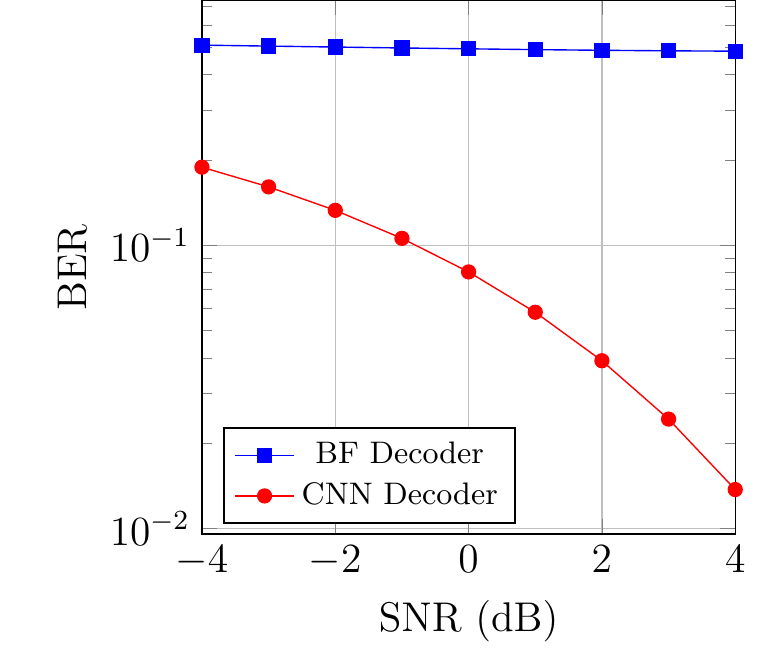} 
    \caption{BER Vs SNR ($n$=1000)}  
    \label{fig:ldpc500ber}
    \end{subfigure}
     ~
    \begin{subfigure}{0.32\textwidth}
    \centering
    \includegraphics[width=2.35in, height=1.72in]{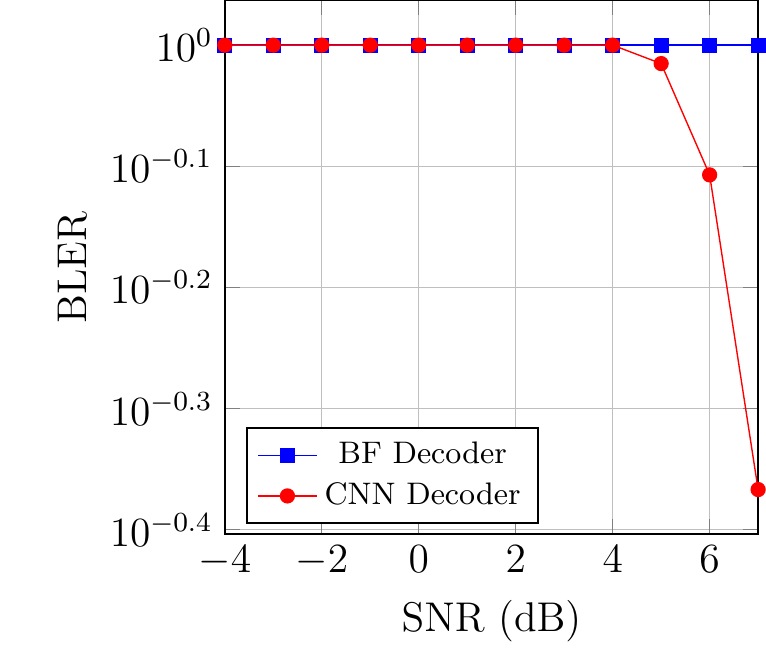} 
    \caption{BLER Vs SNR ($n$=1000)}  
    \label{fig:ldpc500bler}
    \end{subfigure}
\caption{\label{fig:ldpcEncoder} BER and BLER comparison of Bit flipping based decoder and CNN decoder 
}
\end{figure*}

\begin{figure}
\begin{subfigure}[t!]{\columnwidth}
\begin{minipage}{0.48\columnwidth}
\includegraphics[width=1.68in]{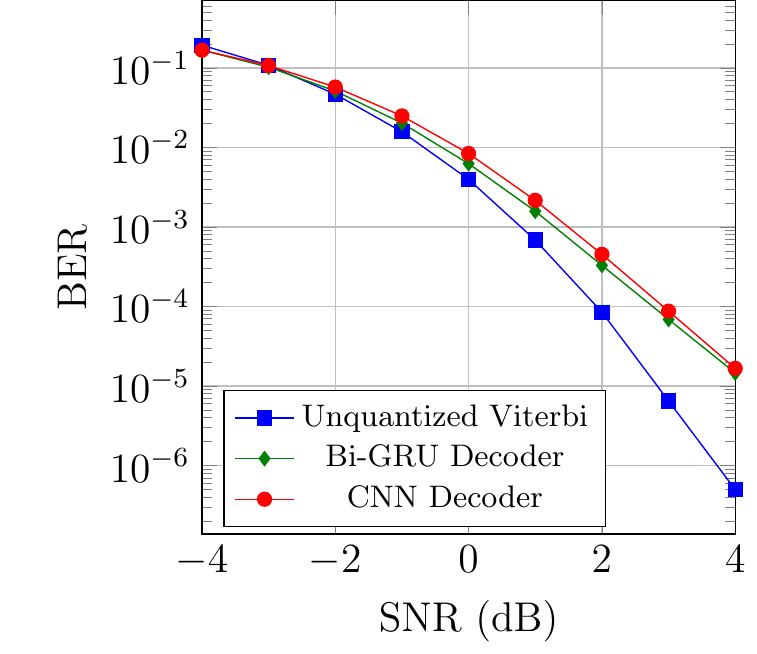}
\caption{BER Vs SNR (Conv. code)}
\label{fig:soft_conv_ber}
\end{minipage}
\begin{minipage}{0.48\columnwidth}
\includegraphics[width=1.68in]{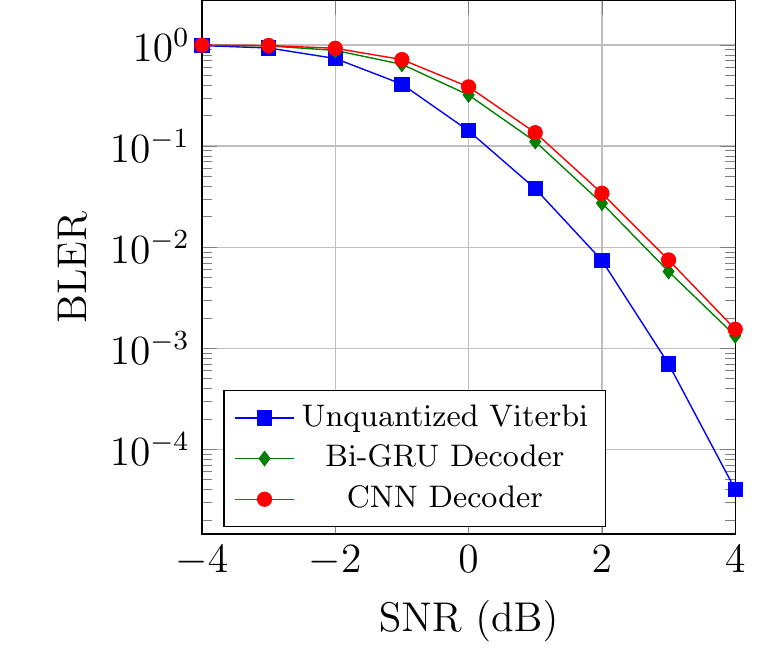}
\caption{BLER Vs SNR (Conv. code)}
\label{fig:soft_conv_bler}
\end{minipage}\\
\begin{minipage}{0.48\columnwidth}
\includegraphics[width=1.68in]{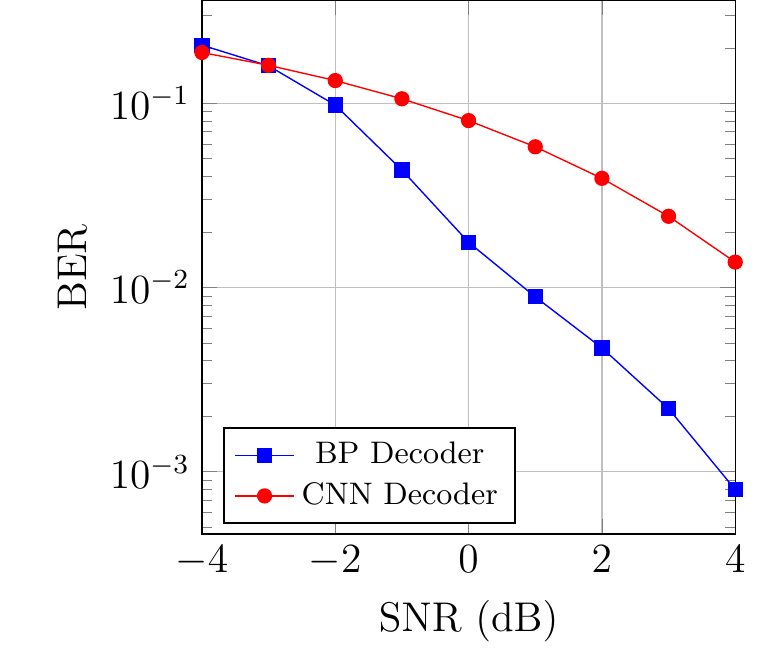}
\caption{BER Vs SNR (LDPC code)}
\label{fig:ldpcbersoft}
\end{minipage}
\begin{minipage}{0.48\columnwidth}
\includegraphics[width=1.68in]{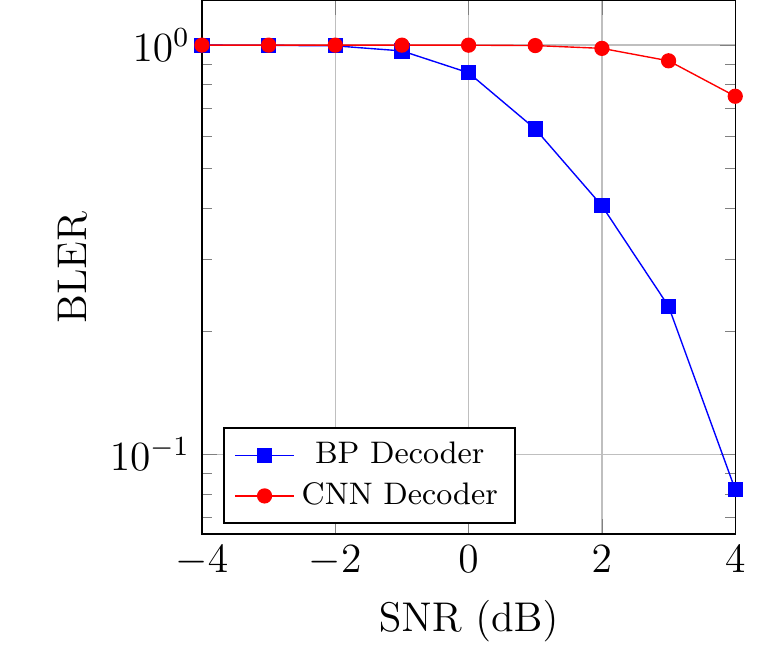}
\caption{BLER Vs SNR (LDPC code)}
\label{fig:ldpcblersoft}
\end{minipage}
\end{subfigure}
\caption{\label{fig:soft_conv} Comparison of CNN decoders with the optimal decoding techniques for convolutional code (figures~(a) and~(b)) and LDPC code (figures~(c) and~(d)). Here, $n = 200$. }
\end{figure}

\begin{figure}[!h]
\begin{subfigure}[t!]{\columnwidth}
\begin{minipage}{0.48\columnwidth}
\includegraphics[width=1.68in]{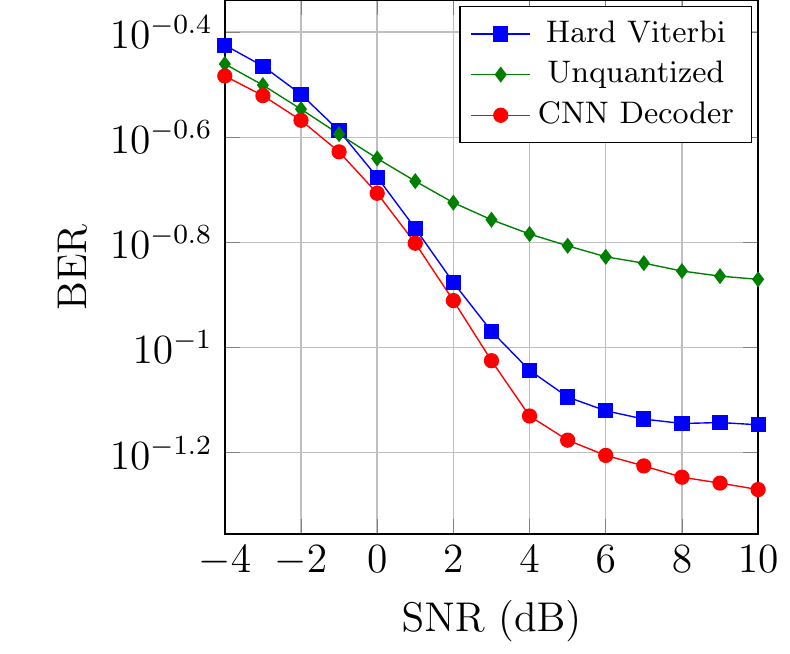}
\caption{$\alpha=0.3$ (Conv. code) }
\label{fig:jerkyconv_30}
\end{minipage}
\begin{minipage}{0.48\columnwidth}
\includegraphics[width=1.68in]{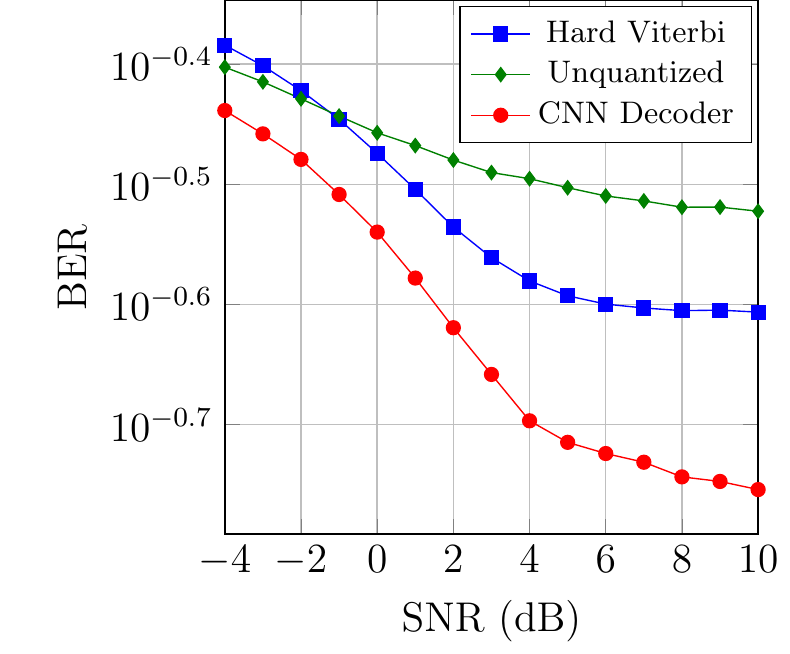}
\caption{$\alpha=0.5$ (Conv. code)}
\label{fig:jerkyconv_50}
\end{minipage} \\
\begin{minipage}{0.48\columnwidth}
\includegraphics[width=1.68in]{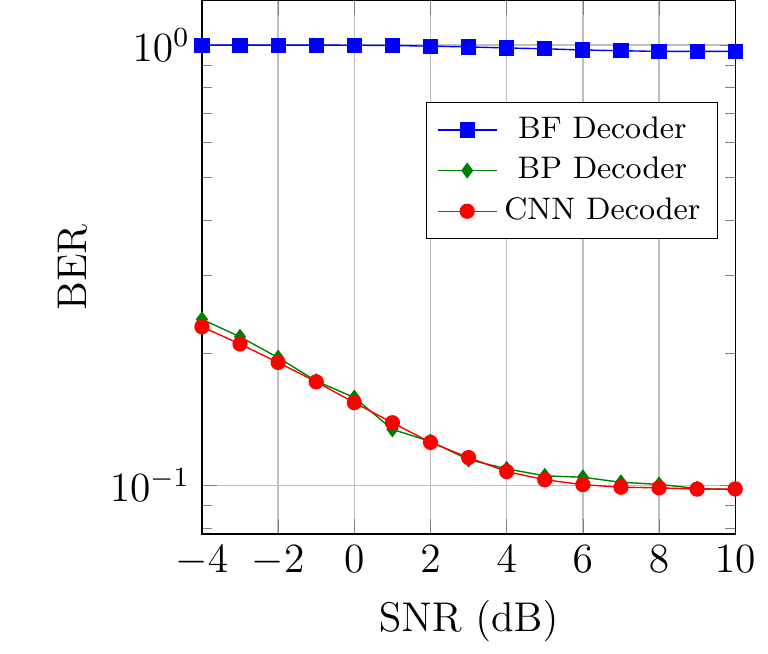}
\caption{$\alpha=0.3$ (LDPC code)}
\label{fig:jerkyldpc_30}
\end{minipage}
\begin{minipage}{0.48\columnwidth}
\includegraphics[width=1.68in]{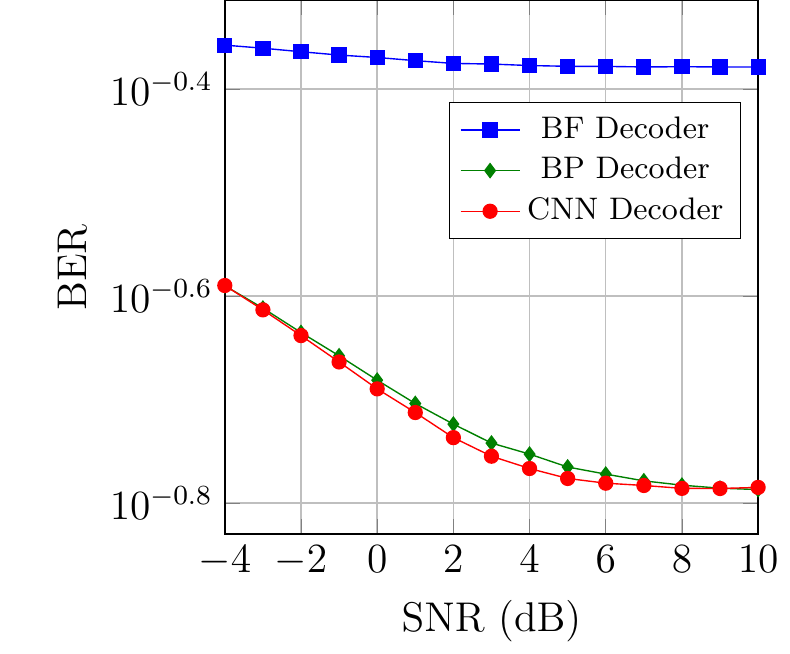}
\caption{$\alpha=0.5$ (LDPC code)}
\label{fig:jerkyldpc_50}
\end{minipage}
\end{subfigure}
\caption{\label{fig:jerky} BER Vs SNR comparison of CNN decoders with the existing decoding techniques for channel with random outage. Here, $n = 200$. (a)~and~(b)~show BER comparison of three convolutional code decoders, viz. CNN decoder, quantized and unquantized Viterbi decoders for outage probabilities $\alpha$ as $0.3$ and $0.5$, resp. (c)~and~(d)~show BER comparison of three LDPC decoders, viz. CNN decoder, bit-flipping and log simple belief propagation decoders with $\alpha$ as $0.3$ and $0.5$, resp. }
\end{figure}

\vspace{-0.1cm}
\subsection{Decoding of LDPC Codes}
 We consider a rate $1/2$ regular LDPC~$(10,20)$ code here. 
 The received noisy signal is used against the original message to train the decoder. The same network architecture described in Table~\ref{table:network} is used for training purpose. The decoding performance is compared against the bit flipping (BF) method for LDPC codes. In Figure~\ref{fig:ldpcEncoder}, performance comparisons are given for block lengths of $100$, $200$ and $1000$. For both BER and BLER, the CNN decoder outperforms the BF decoder. 

To further evaluate the decoder performance, we now compare the CNN decoder with soft decoding schemes.

\vspace{-0.1cm}
\subsection{Comparison with soft decoding schemes}
For comparing the performance on decoding convolutional codes, the unquantized Viterbi decoder is used as a reference. This is presented in~Fig.~\ref{fig:soft_conv_ber} and~\ref{fig:soft_conv_bler}. 
Although the CNN decoder outperforms the hard Viterbi decoder, it still has got some way to go in reaching the optimal unquantized Viterbi decoder. For LDPC codes, the
belief propagation (BP) decoder performs better than the CNN decoder in terms of both BER and BLER as shown in Fig.~\ref{fig:ldpcbersoft} and~\ref{fig:ldpcblersoft}. Thus, more
work is required for any existing learning based method, be it CNN or RNN, to perform at the level of any near-optimal soft decoding scheme. However, this situation dramatically changes when the channel statistics are not exactly known. In particular, for models having
channel variations which are not exactly tracked at the receiver, the optimal decoder design is much more complex. An optimal decoder designed for a particular channel law may fare poorly due to the
\emph{mismatched} decoding when the encountered channel is different. However, a properly trained CNN scheme can still
offer reliability, which we now demonstrate by experiments.


\subsection{Random channel outages}
In  systems like mmWave 5G links, random channel outages can 
occur. During an outage, the channel SNR suddenly drops to a lower value before getting restored to its proper level. This results in different noise distributions within the same code block, a condition akin to arbitrary varying channels in information theory~\cite{el2011network}.
In 5G systems, the  SNR difference can get as high as $20$dB~\cite{rappaport2013millimeter}. In order to show the robustness of the CNN decoder, a channel where each symbol independently experiences outage with probability~$\alpha$ is considered.  The SNR during an outage is taken as $-10$dB. We compare the performance of an appropriately trained CNN decoder with hard and unquantized Viterbi decoder for $\alpha=0.3$ and $\alpha=0.5$, the results are presented in Fig.~\ref{fig:jerkyconv_30} and~\ref{fig:jerkyconv_50}  respectively. 
Clearly the CNN decoder outperforms
the Viterbi decoder for both the cases.
Notice that the unquantized Viterbi decoder is optimal for a fixed channel, but becomes mismatched in presence of unknown channel quality changes.
The same experiment is now repeated for LDPC codes. Fig.~\ref{fig:jerkyldpc_30} and~\ref{fig:jerkyldpc_50} compare the CNN decoder with bit flipping as well as belief propagation methods for $\alpha=0.3$ and $\alpha=0.5$. For this case, CNN decoder outperforms the bit flipping method. It outperforms the belief propagation method when $\alpha=0.5$ and gives comparable performance for $\alpha=0.3$.

This shows that the CNN decoder not only  is robust, but also can  outperform analytically derived decoders when channel conditions are varying randomly. The application of CNN decoder can become even more important with real world systems where the channel model is not  analytically available. We leave this application as a future work.

\section{Conclusion}\label{sec:conclusion}
In this paper, we proposed a novel training strategy called MIST, and showed that CNN based neural net decoders can be trained for convolutional and LDPC codes. The training method is scalable to higher block lengths. The  proposed decoder matches the state-of-the-art RNN based decoder in their performance while providing significantly lower decoding latency. Furthermore, when the channel suffers random outages, the neural net based decoders outperform even the analytically derived decoders which are optimal for a fixed channel law. 

Designing decoders for arbitrary varying channels is left as a future work. Improving the CNN architecture to match soft decoding schemes for fixed channels is another open problem of interest.


\bibliographystyle{IEEEtran}      
\bibliography{myreferences}

\end{document}